# Detection of poloidal magnetic flux emission from a plasma focus - First Experiments at Sofia University

A B Blagoev[1,2] and S K H Auluck[2]

**Abstract.** The existence of axial (poloidal) magnetic field in a plasma focus and its significance in plasma focus phenomenology has been extensively discussed in a recent review paper. The poloidal magnetic field is a part of the transient 3-dimensional magnetic field structures which arise spontaneously, accelerate ions and keep them moving in trajectories that repeatedly cross a dense and warm plasma target. This is the origin of the abnormally high fusion reaction rate of the plasma focus, which has been known since the 1960's but has begun to be understood quite recently. Further progress now depends on explorations of the global aspects of the evolution of poloidal magnetic field. However, well-known experimental difficulties involved in standard techniques of axial magnetic field measurement hamper such research efforts. Taking cognizance of this stalemate, the International Scientific Committee for Dense Magnetized Plasmas launched an initiative to address this state of affairs in an International Video Conference on 24th April 2020. This paper reports on the first experiments that resulted from that initiative.

## 1. Introduction

The plasma focus is well known [1,2,3] as a prolific source of d-d fusion neutrons since the 1960s. The SPEED-II plasma focus reportedly produced $10^{12}$ d-d neutrons in a few exceptional shots [4] with an energy input of only 80 kJ. Considering that each neutron producing reaction was accompanied with an energy output of 3.27 MeV in the neutron branch and 4.03 MeV in the proton branch, the d-d fusion output was ~1 J in these exceptional shots. This level of wall-plug efficiency has not been reported from any fusion reactor candidate even till now.

Still, the plasma focus is not considered as a fusion reactor candidate This is because it was decided in the early days of Controlled Fusion Research that it is impossible to scale up a fusion device to the level of net energy gain unless the fusion reaction occurred between the high energy tail of a Maxwellian velocity distribution of ions and the majority of moderate energy ions in a hot, confined plasma. This issue was deemed to be permanently settled, no longer open to serious scientific scrutiny or debate. This Fundamental Premise of Controlled Fusion Research eliminated the plasma focus from the race for Controlled Fusion since it was conclusively proved that these abundant neutrons were anisotropically produced by ions having energy ~ 100 keV, much above the estimated electron temperature of the plasma.

[1] Faculty of Physics at the University of Sofia, 5 J. Bourchier blv. 1164, Sofia, .Bulgaria. blagoev@phys.uni-sofia.bg

[2] International Scientific Committee on Dense Magnetized Plasmas, Hery 23, P.O. Box 49, 00-908 Warsaw, Poland, skhauluck@gmail.com

Recent advances in plasma focus research have raised a serious question mark [3] over this Fundamental Premise of Controlled Fusion Research (FPCFR). Experimental data from many plasma focus devices cross-correlated across multiple diagnostics is shown to be consistent with the following ICDMP BEAM-TARGET Model.

The fusion reactions are correlated in both time and space with the spontaneous formation and decay of 3-dimensional (3-D) bounded plasma structures visible in interferometry. Their boundaries can be maintained over the time of observation only if there exist 3-D distributions of magnetic fields to confine them. Ions are accelerated by the associated transient 3-D electric fields and trapped in trajectories that repeatedly cross the dense plasma [5]. Such scenarios, where ions are accelerated and confined to a finite volume and where they repeatedly traverse a bounded warm plasma target over a folded trajectory much longer than dimensions of the plasma, was inconceivable in the early days of Controlled Fusion Research when the Fundamental Premise of Controlled Fusion Research (FPCFR) came to be tacitly adopted as a self-evident truth. The extensive evidence-base of this model [3] suggests that the time is now ripe for a push towards re-examination of the FPCFR. This evidence-base includes direct measurement of axial magnetic field both in the radial implosion phase and in the pinch phase using specially constructed magnetic probes and reaction product images and spectra which exhibit unmistakable signatures of this mechanism.

In spite of such direct evidence, a substantial amount of the evidence that underlies the ICDMP BEAM-TARGET Model is indirect and requires some degree of reasonable interpretation which is discussed extensively [3]. Additionally, such evidence needs to be collected from a wide spectrum of plasma focus devices, small, medium and large to establish its universality. Since the question of elevation of the plasma focus to the status of a fusion reactor candidate is likely to be revisited sooner or later, it would be desirable to obtain direct supporting evidence for this model. It would be particularly important to understand the phenomenology of spontaneous generation of 3-D magnetic fields.

The axial (poloidal) magnetic field is the least understood and investigated component of such phenomenology because traditional methods of its measurement suffer from significant experimental difficulties. The magnetic probe method is invasive and suffers from contamination from the capacitive coupling with the plasma and from the magnetic field components other than the one being measured. Faraday rotation technique yields the product of electron density and magnetic field integrated along the path of the laser beam which is not of sufficient value in terms of interpretation to justify the effort and expense involved. Further exploration of the phenomenology of spontaneous generation of 3-D magnetic fields and its correlation with fusion reactions thus remains in jeopardy.

The International Scientific Committee for Dense Magnetized Plasmas organized an International Video Conference [6] on April 24, 2020 to launch an initiative to address such stalemate in this crucially important topic of research. Experimental efforts were immediately started at the University of Sofia to implement a proposed scheme for detection of the poloidal magnetic field suggested earlier [5]. First results from this scheme were presented at the $9^{th}$ ISSWPP held in December 2020 [6]. Lessons learnt were used to refine the technique proposed in the video conference [7] and are now published as a tutorial paper [8]. The first experimental results using the improved technique were presented at the $10^{th}$ ISSWPP in July 2022 and are being formally reported in this paper.

This paper is organized as follows. Section 2 briefly reviews the technique and brings out the difference between the new technique and its earlier version. Section 3 describes the experimental apparatus. Section 4 presents the results and their main features. Section 5 concludes the paper with a brief summary.

**2. Brief overview of the experimental technique**
It is important to appreciate that the objective of exploring the global evolution of the 3-D magnetic structure in correlation with other standard diagnostics cannot be achieved by a local measurement of magnetic field at a point or by mapping a spatial distribution at an instant in time, although both approaches certainly provide useful information. Instead, a signal whose amplitude and temporal profile is related to the entire 3-D magnetic structure would be more useful. Such a signal is found in the poloidal magnetic flux (PMF) emission. Any 3-D magnetic field distribution would have a poloidal magnetic flux that "returns" outside the current (or magnetic moment) distribution that creates it. This is a consequence of conservation of magnetic flux either in infinite space or within a continuous metallic boundary such as a vacuum chamber. A portion of this magnetic flux can be sampled by a detector that is outside this current distribution (the plasma focus electrode structure in the present case). The signal from such detector would arise synchronously with the 3-D magnetic structure and would be proportional to the current associated with the 3-D structure. A pure z-pinch with only axial current would not be able to create a poloidal magnetic flux (PMF). The time variation of the PMF signal could be studied in correlation with other standard diagnostics to decipher its significance in plasma focus phenomenology.

One version of this diagnostics [5] consists of two single conducting loops placed outside the cathode of the plasma focus and connected in clockwise (CW) and counter-clockwise (CCW) sense to the diagnostic cables. The two signals are seen to be different [6] because the capacitive coupling between the plasma and the conducting single loops does not depend on their orientation but the PMF signal reverses its sign. Half the sum of these signals then represents the capacitive coupling and half the difference represents the PMF signal. Theory of this measurement [5] reveals that the SUM signal is a common mode signal: both the inner and outer conductor of the cable are elevated in potential with respect to the oscilloscope ground. This is potentially problematic since the common mode signal can involve the ground loop circuit of the facility unless the oscilloscope is operated with a battery-based inverter power supply that is isolated from the laboratory ground. This is not a desirable technical solution. Another problem is that the area through which the PMF passes includes the outer boundary of the plasma which is moving so that the PMF is not defined over a fixed detector dimension.

Both these deficiencies are corrected in the new diagnostic, schematically shown in Fig 1. Its theory of measurements has been presented in a tutorial format [7] along with a procedure for its implementation and testing.

The detector consists of two identical double loops wrapped around a plastic tube centred with respect to the axis of the plasma focus, one where the inner loop is oriented in the $+\theta$ (CCW) direction and the other in the $-\theta$ (CW) direction. The inner loop is connected to the centre conductor of the 50 Ohm cable for both. The outer loop is electrostatically shielded from the capacitive coupling from the plasma and hence its potential is governed by its connection with the oscilloscope ground, eliminating the common-mode signal encountered in the single loop case [5,6]. Secondly, the PMF signal can arise only from the returning PMF over the annular space in the double-loop structure, which is independent of plasma motion.

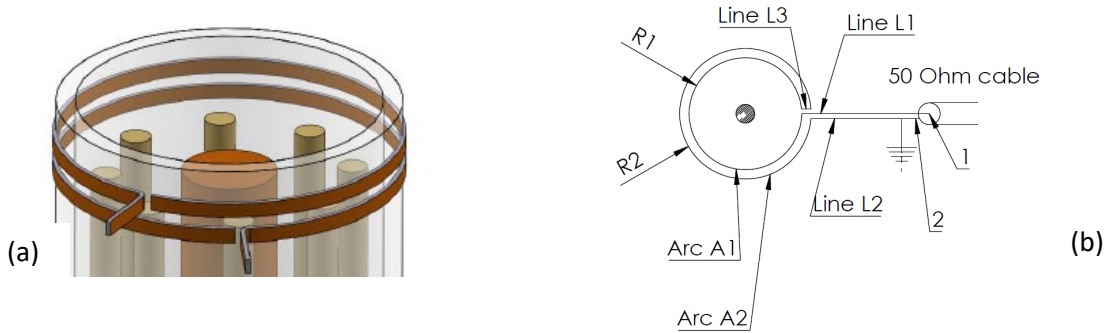

**Figure 1.** Schematic illustration of the PMF emission detector. A plastic tube placed outside the cathode and centred with respect to the axis supports two shorted microstrip transmission lines wrapped in clockwise and counter-clockwise directions, with the inward facing conductor connected to the inner cable conductor.

The signals from the two double-loops are then given by

$$V_{\text{CCW}} = R_Z \frac{\text{d}(C_1 V_0)}{\text{d}t} + V_{\text{PMF}} \frac{\text{d}\phi_{\text{PMF}}}{\text{d}t}; V_{\text{CW}} = R_Z \frac{\text{d}(C_1 V_0)}{\text{d}t} - V_{\text{PMF}}; V_{\text{PMF}} \equiv -\frac{\text{d}\phi}{\text{d}t} \qquad 1$$

Making symmetric and anti-symmetric combinations of the two signals,

$$V_{\text{SYM}} \equiv \frac{1}{2}(V_{\text{CCW}} + V_{\text{CW}}) = R_Z \frac{d(C_1 V_0)}{dt}; V_{\text{ASYM}} \equiv \frac{1}{2}(V_{\text{CCW}} - V_{\text{CW}}) = V_{\text{PMF}} \qquad 2$$

Here, $C_1$ is the effective capacitance between the plasma and the inner loop, $V_0$ is the effective potential of the plasma with respect to the laboratory (oscilloscope) ground and $R_Z$ is the cable impedance.

The plasma has a net charge density that is given by the divergence of the electric field arising from the motion of the plasma in its magnetic field:

$$\rho = \varepsilon_0 \vec{\nabla} \cdot \vec{E} = -\varepsilon_0 \vec{\nabla} \cdot (\vec{v} \times \vec{B}) \qquad 3$$

The signal $V_{\text{SYM}}$ is then proportional to the rate of change of the right-hand side of equation 3.

One of the important points brought out in the theory of measurement [7] is that both the SYM and ASYM signals are insensitive to azimuthal variations in the plasma properties.

## 3. Experimental apparatus:

This experiment was performed on the University of Sofia plasma focus [6]. This 3kJ Mather type PF device has a total capacitance of the condenser bank ~ 20 µF, with maximal charging voltage of 40 kV. The main switch of the discharge circuit is a vacuum spark gap. The anode is a hollow copper tube with 2 cm diameter and 14.5 cm length. The cathode consists of six copper rods (0.8 cm diameter, 16 cm length) mounted in the massive cathode base on a circle with a 3.5 cm radius. The vacuum chamber has 15.5 cm inner diameter and 35 cm height. The working gas can be air, Ar or Deuterium. In this investigation Ar was used. The voltage operating range in the case of air or Ar is 12÷18 kV and the optimal pressure is in the range of 0.8÷1.3 mTorr).

Diagnostics include the discharge current (pick-up coil), the current derivative (Rogowski belt), the soft X-ray (PIN diodes) and the hard X-ray emission (scintillator probe) from the plasma. Thermo-luminescent dosimeters (TLD) are used to measure the full X-ray dose in the chamber of the PF and to control the radiation outside the device. The data of all of the detectors mentioned above have been recorded by two 4 channel Tektronix oscilloscopes (TDS 3034C, TDS 2004B).

The coils, shown on Figure 1a, are made from a 12 mm wide copper tape (type FE-5100-5299-9 3M). Both the CW and CCW coils are closely situated to each other and fixed on a 110 mm wide and 50 mm long PVC tube. The position of the coils was chosen so that the plane between them passes through the midplane where the pinches occur.

The double loops were made by folding the copper adhesive tape along with its insulating liner fixing them on the support (110 mm diameter PVC tube of 50 mm height). At 35 µm thickness of the metal layer, the distance between the two metal layers is about 270 microns including the width of the two protective plastic layers of the tape. Thus, a shorted microstrip transmission line is effectively formed. The width of this sandwich was adjusted in such manner as to match its theoretical impedance with the 47 Ohm coaxial cable used for measurement according to a standard procedure [25]. Possibility of frequency response distortion because of the variation of the dielectric constant of the plastic layers and adhesive foam with frequency was examined by conducting a square-pulse response test of the double loops connected with the cable [25], which was quite satisfactory. Since the aim of the experiment was detection of poloidal magnetic flux emission – just a yes/no answer – no further characterization of this diagnostic was performed, leaving that to a future campaign.

## 4. Experimental results:

First experiments at the Sofia University plasma focus had a limited aim of demonstrating that this newly proposed technique actually works and produces interpretable results. This section presents some illustrative examples towards that aim.

The first example is of Shot #131 at charging voltage U=15.0 kV, gas pressure p= 1.05 Torr Ar

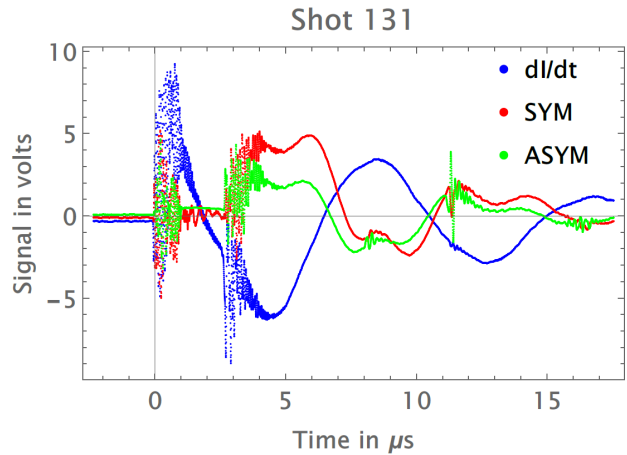

**Figure 2.** Shot #131

The standard measurement of the current derivative dI/dt has a characteristic sharp dip that marks the time of collapse of the imploding plasma on the axis. The subsequent plasma dynamics has no influence on this signal. In contrast, the SYM and ASYM signals show activity that extends beyond this event but relatively less activity before. The rate of change of poloidal magnetic flux (ASYM signal) increases abruptly after the plasma collapse on the axis. This indicates activation of a dynamo process that creates a poloidal component of magnetic field. Simultaneous and similar increase in the SYM signal shows that this process is related to the post-pinch expansion of the plasma that is known to produce a plasma bubble that expands both radially and axially. This corroborates the interpretation of PF-1000 phenomenology in Section 4.3.1.3 of the review paper [3]. The continuation of significant activity in both the SYM and ASYM signals much after the pinch phase corroborates the observations of significant post-pinch activity discovered by Neil and Post [9] by $CO_2$ laser scattering.

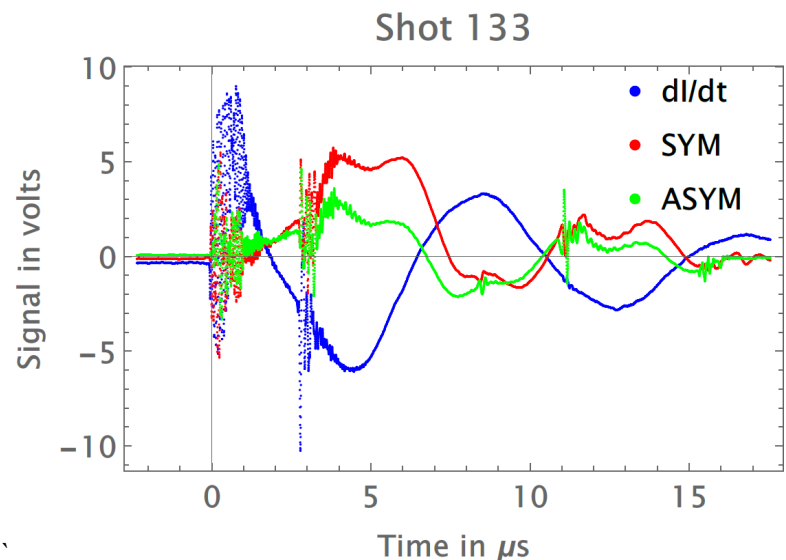

**Figure 3.** Shot #133

The next example is of Shot #133, at U=14.5 kV and p=1.08 Torr Ar (Fig 3). Both the SYM and ASYM signals have significant activity before the plasma collapse on the axis. This is in conformity with the PF-1000 magnetic probe data [3] which reveals the existence of poloidal magnetic field in the radial implosion phase. Both signals undergo a rapid change along with the current derivative signal.

This indicates occurrence of a rapid topological change in the magneto-plasma structure during the pinch phase. Such rapid changes in topology are observed by 15-frame interferometry on the One Megajoule Plasma Focus PF-1000 and found to be correlated with neutron emission [3]. The linear rise in both the SYM and ASYM signals before the current derivative singularity indicate presence of a dynamo mechanism that is amplifying the poloidal magnetic field.

These two examples are sufficient to conclude that the new diagnostics for detection of poloidal magnetic flux emission from a plasma focus actually works and provides information relevant to plasma focus phenomenology with rather modest resources and effort. Further discussion of this data will be reported in a more specialized forum.

## 5. Summary and conclusions:

This paper formally reports results from the first experiments at the Sofia University plasma focus aimed at demonstrating the feasibility and utility of a newly proposed diagnostic for detection of poloidal magnetic flux emission from a plasma focus. This diagnostic provides two signals, one which is proportional to the rate of change of charge density $\rho = -\varepsilon_0 \vec{\nabla} \cdot (\vec{v} \times \vec{B})$ and the other is equal to the rate of change of poloidal magnetic flux through an annular gap in the detector.

First results indicate that these signals have features that corroborate the Emerging Narrative of plasma focus phenomenology inferred from a wide spectrum of data from many plasma focus devices cross-correlated across multiple diagnostics. Significant rapid changes in these signals coinciding with the current-derivative singularity and subsequent rise are consistent with rapid topological changes observed using 15 frame interferometry on PF-1000 and growth, persistence and decay of spontaneously generated 3-D bounded structures.

In view of its novelty, usefulness and modest resource requirement, it is hoped that this technique will be used in correlation with standard diagnostics on multiple plasma focus devices.

**Acknowledgments**
Authors wish to gratefully acknowledge significant assistance from Mladen Mitov, Stanislav Zapryanov and Vasil Yordanov.
This work is partially supported by the Grant № BG05M2OP001-1.002-0019:
"Clean Technologies for Sustainable Environment - Waters, Waste, Energy for a Circular Economy", financed by the Science and Education for Smart Growth Operational Program (2014-2020)
and co-financed by the EU through the ESIF